\documentstyle[11pt,epsfig,psfig]{article}

\begin{document}

\begin{center}
{\bf Pion distribution amplitudes within the instanton model of QCD vacuum}
\\[0pt]
\vspace{3mm} {\small A. E. Dorokhov} \\[0pt]
\vspace{1mm} {\small {\it Bogoliubov Laboratory of Theoretical Physics,
Joint Institute for Nuclear Research, \\[0pt]
141980 Dubna, Russia}}
\end{center}

\abstract{
{\small
Pion transition form factor for the process $\gamma ^{\star }\gamma
^{\star}\rightarrow \pi ^{0}$ at space-like values of photon momenta
is calculated within the effective quark-meson model with the interaction
induced by instanton exchange.
The leading and next-to-leading order power asymptotics of the form factor and
the relation between the light-cone pion distribution
amplitudes of twists 2 and 4 and the dynamically generated quark mass are found.}}
\\[0.5cm]

The pion form factor $M_{\pi ^{0}}(q_{1}^{2},q_{2}^{2})$ for the transition
process $\gamma ^{\star }(q_{1})\gamma ^{\star }(q_{2})\rightarrow \pi
^{0}(p)$, where $q_{1}$ and $q_{2}$ are photon momenta, is related to
fundamental properties of QCD dynamics at low and high energies. At zero
photon virtualities the observed value of the width for the two-photon decay
of the $\pi _{0}-$meson
\begin{equation}
\Gamma (\pi ^{0}\rightarrow \gamma \gamma )=\frac{e^{2}m_{\pi _{0}}^{3}}{%
64\pi }M_{\pi ^{0}}^{2}\left( 0,0\right) =7.79(56)\ {\rm eV,}
\end{equation}%
is consistent with the theoretical prediction due to the chiral anomaly for $%
\pi _{0}$
\begin{equation}
M_{\pi ^{0}}\left( 0,0\right) =(4\pi ^{2}f_{\pi })^{-1},
\end{equation}%
where $f_{\pi }=92.4$ {\rm MeV }is the pion weak decay constant{\rm .}

The existing experimental data from CELLO \cite{CELLO} and CLEO \cite{CLEO}
Collaborations on the form factor $M_{\pi ^{0}}$ for one photon being almost
real, $q_{2}^{2}\approx 0$, with the virtuality of the other photon scanned
up to $8$ GeV$^{2}$, can be fitted by a monopole form factor:
\begin{equation}
\left. M_{\pi ^{0}}(q_{1}^{2}=-Q^{2},q_{2}^{2}=0)\right\vert _{fit}=\frac{%
g_{\pi \gamma \gamma }}{1+Q^{2}/\Lambda _{\pi }^{2}},\ \ \ \ \ \Lambda _{\pi
}\simeq 0.77\ GeV,  \label{Fpiggfit}
\end{equation}%
where $g_{\pi \gamma \gamma }=0.275$ {\rm GeV}$^{-1}$ \ is the two-photon
pion decay constant. The large $Q^{2}$ behavior of the form factor (\ref%
{Fpiggfit}) is in agreement with the lowest order perturbative QCD~(pQCD)
prediction \cite{BrLep79}
\begin{equation}
\left. M_{\pi ^{0}}(q_{1}^{2},q_{2}^{2})\right\vert _{Q^{2}\rightarrow
\infty }=J^{(2)}\left( \omega \right) \frac{1}{Q^{2}}+J^{(4)}\left( \omega
\right) \frac{1}{Q^{4}}+O(\frac{\alpha _{s}}{\pi })+O(\frac{1}{Q^{6}}),
\label{AmplAsympt}
\end{equation}%
where the leading (LO) and next-to-leading (NLO) order asymptotic
coefficients $J\left( \omega \right) $ are expressed in terms of the
light-cone pion distribution amplitudes (DA), $\varphi _{\pi }(x)$:
\begin{equation}
J^{(2)}\left( \omega \right) =\frac{4}{3}f_{\pi }\int_{0}^{1}dx\frac{\varphi
_{\pi }^{(2)}(x)}{1-\omega ^{2}(2x-1)^{2}},\ \ \ \ J^{(4)}\left( \omega
\right) =\frac{4}{3}f_{\pi }\Delta ^{2}\int_{0}^{1}dx\frac{1+\omega
^{2}(2x-1)^{2}}{[1-\omega ^{2}(2x-1)^{2}]^{2}}\varphi _{\pi }^{(4)}(x).
\label{J}
\end{equation}%
In the above expressions $Q^{2}=-(q_{1}^{2}+q_{2}^{2})\geq 0$ is the total
virtuality of photons and
$\omega=(q_{1}^{2}-q_{2}^{2})/(q_{1}^{2}+q_{2}^{2})$ is the asymmetry in their
distribution. The distribution amplitudes are normalized as $%
\int_{0}^{1}dx\varphi _{\pi }(x)=1$ and the parameter $\Delta ^{2}$
characterizes the scale of the NLO power corrections. The first perturbative
correction to the LO term in (\ref{AmplAsympt}) has been found in \cite%
{Braaten} and the NLO power corrections have been discussed in \cite{svzB237},
\cite{Gorsky:nm} and more recently in \cite{twist4} within the light-cone sum rules.

The leading momentum power dependence of the form factor (\ref{AmplAsympt})
is dictated by the scaling property of the pion DA. But the
coefficients of the power expansion depend crucially on the internal
pion dynamics, which is parameterized by the nonperturbative pion DAs, $%
\varphi _{\pi }(x)$, defined at some normalization scale $\mu $, with $x$
being the fraction of the pion momentum, $p$, carried by a quark. At
asymptotically large normalization scale $\mu \rightarrow \infty $ the DAs
are determined in pQCD:
\begin{equation}
\varphi _{\pi ,as}^{(2)}(x)=6x(1-x),\qquad \varphi _{\pi
,as}^{(4)}(x)=30x^{2}(1-x)^{2}.  \label{DAas}
\end{equation}%
However, for the description of the experimentally observable hard exclusive
processes one needs to know the DAs normalized at virtuality $\mu ^{2}\sim 1$
GeV$^{2}$. The aim of this letter is to calculate the pion transition form
factor in the kinematical region up to moderately large $Q^{2}$ and extract
from its power expansion in $1/Q^2$ the pion DAs at normalization scale typical for
hadrons. The calculations carried out within the effective model with
nonlocal quark-quark interaction are consistent with the chiral anomaly and
result in the relations between the DAs of twists 2 and 4 and the
dynamically generated nonlocal quark mass. The usage of the covariant
nonlocal low-energy model based on the Schwinger-Dyson approach to dynamics
of quarks and gluons has many attractive features as the approach preserves
the gauge invariance, it is consistent with the low-energy theorems and
takes into account the large distance dynamics of the bound state.
Furthermore, the intrinsic nonlocal structure of the model may be motivated
by fundamental QCD interactions induced by the instanton and gluon exchanges.

The effective quark-pion dynamics motivated by the instanton-induced
interaction\footnote{%
See for a review {\it e.g.} \cite{ADoLT00}.} may be summarized in terms of
the dressed quark propagator
\[
S^{-1}\left( p\right) =\widehat{p}-M\left( p^{2}\right) ,
\]%
the quark-pion vertex
\[
\Gamma _{\pi }^{a}\left( k,p,k^{\prime }=k+p\right) =\frac{i}{f_{\pi }}%
F(k^2,k^{\prime 2})\gamma _{5}\tau ^{a},\qquad F\left( k^2,k^{\prime
2}\right) =\sqrt{M\left( k^{2}\right) M\left( k^{\prime 2}\right) },
\]%
and the quark-photon vertex satisfying the Ward-Takahashi identity%
\[
\Gamma ^{\mu }\left( k,q,k^{\prime }=k-q\right) =eQ\left[ \gamma _{\mu
}-\left( k+k^{\prime }\right) _{\mu }G\left( k^2,k^{\prime 2}\right) \right]
,\qquad G\left( k^2,k^{\prime 2}\right) =\frac{M\left( k^{\prime 2}\right)
-M\left( k^{2}\right) }{k^{\prime 2}-k^{2}},
\]%
where $M(k^{2})$ is the dynamically generated quark mass. The dynamical
quark mass characterizes the momentum dependence of an order parameter for
spontaneous breaking of the chiral symmetry and may be expressed in terms of
the gauge invariant nonlocal quark condensate \cite{ADWB01}. The inverse
size of the nonlocality scale, $\Lambda $, is naturally related to the
average virtuality of quarks that flow through the vacuum, $\lambda
_{q}^{2}\sim \Lambda ^{2}$. The value of $\lambda _{q}^{2}$ is known from
the QCD sum rule analysis, $\lambda _{q}^{2}$ $\approx 0.4\pm 0.1{\rm GeV}%
^{2}$ \cite{BI82}, and, within the instanton model, may be expressed through
the average instanton size, $\rho _{c}$, as $\lambda _{q}^{2}\approx 2\rho
_{c}^{-2}$ \cite{DEM97}.
The pion weak decay constant is expressed by the Pagels-Stokar formula
\begin{equation}
f_{\pi }^{2}=\frac{N_{c}}{4\pi ^{2}}\int_{0}^{\infty }du\frac{uM(u)\left[
M(u)-uM^{\prime }(u)/2\right] }{D^{2}(u)},  \label{f_pi}
\end{equation}%
where $M^{\prime }(u)=\frac{d}{du}M(u)$ and $D(u)=u+M^{2}(u)$.

The invariant amplitude for the process $\gamma ^{\ast }\gamma ^{\ast
}\rightarrow \pi ^{0}$ is given by
\[
A\left( \gamma ^{\ast }\left( q_{1},\epsilon _{1}\right) \gamma ^{\ast
}\left( q_{2},\epsilon _{2}\right) \rightarrow \pi ^{0}\left( p\right)
\right) = -ie^2\varepsilon _{\mu \nu \rho \sigma }\epsilon _{1}^{\mu}
\epsilon _{2}^{\nu }q_{1}^{\rho }q_{2}^{\sigma }M_{\pi ^{0}}\left(
q_{1}^{2},q_{2}^{2}\right) ,
\]%
where $\epsilon _{i}^{\mu }$ are the photon polarization vectors. In
the effective model one finds the contribution of the triangle diagram to
the invariant amplitude as
\begin{equation}
A\left( \gamma _{1}^{\ast }\gamma _{2}^{\ast }\rightarrow \pi ^{0}\right)
=-ie^2\frac{N_{c}}{3f_{\pi }}\int \frac{d^{4}k}{(2\pi )^{4}}%
F(k_{+}^2,k_{-}^2)\left\{ tr[i\gamma _{5}S(k_{-})\widehat{\epsilon }%
_{2}S[k-q/2]\widehat{\epsilon }_{1}S(k_{+})]+\right.  \label{MPiGG}
\end{equation}%
\[
+tr[i\gamma _{5}S(k_{-})S[k-q/2]\widehat{\epsilon }_{1}S(k_{+})]\left(
\epsilon _{2},2k-q_{1}\right) G\left( (k-q/2)^2,k_{-}^2\right)
\]%
\[
\left. +tr[i\gamma _{5}S(k_{-})\widehat{\epsilon }_{2}S[k-q/2]S(k_{+})]%
\left( \epsilon _{1},2k+q_{2}\right) G\left( k_{+}^2,(k-q/2)^2\right)
\right\} +\left( q_{1}\leftrightarrow q_{2};\epsilon _{1}\leftrightarrow
\epsilon _{2}\right) ,
\]%
where $p=q_{1}+q_{2},$ $q=q_{1}-q_{2},$ $k_{\pm }=k\pm p/2$. In the adopted
chiral limit $\left( p^{2}=m_{\pi }^{2}=0\right) $ with both photons real $%
\left( q_{i}^{2}=0\right) $ one finds the result
\begin{equation}
M_{\pi ^{0}}\left( 0,0\right) =\frac{N_{c}}{6\pi ^{2}f_{\pi }}%
\int_{0}^{\infty }du\frac{uM(u)\left[ M(u)-2uM^{\prime }(u)\right] }{D^{3}(u)%
}=\frac{1}{4\pi ^{2}f_{\pi }},  \label{ChAn}
\end{equation}%
consistent with the chiral anomaly.

The LO behavior of the form factor at large photon virtualities is given by
the contribution of the first term in (\ref{MPiGG}) and the NLO power
corrections are generated by the second and third terms in (7) and also
appear as the correction to the first term. Thus, for large $%
q_{1}^{2}=q_{2}^{2}=-Q^{2}/2$ and $p^{2}=0$ the form factor has the
asymptotics
\begin{equation}
\left. M_{\pi ^{0}}\left( -Q^{2}/2,-Q^{2}/2\right) \right\vert
_{Q^{2}\rightarrow \infty }=\frac{4f_{\pi }}{3Q^{2}}\left( 1+\frac{\Delta
^{2}}{Q^{2}}\right) +O(\frac{1}{Q^{6}}),\
\end{equation}%
\begin{equation}
\ \Delta ^{2}=\frac{N_{c}}{4\pi ^{2}f_{\pi }^{2}}\int_{0}^{\infty }du\frac{%
u^{2}M(u)(M(u)+\frac{1}{3}uM^{\prime }(u))}{D^{2}(u)},  \label{PowCorr}
\end{equation}%
which is in agreement with the expressions (\ref{AmplAsympt}), (\ref{J}) for
the asymptotic coefficients at $\omega =0$. The parameter $\Delta ^{2}$ has
an extra power of $u$ in the integral with respect to (\ref{f_pi}) and thus
it is proportional to the matrix element $\left\langle \pi (p)\left\vert
g_{s}\overline{d}\tilde{G}_{\alpha \mu }\gamma _{\alpha }p_{\mu
}u\right\vert 0\right\rangle $. The power correction (\ref{PowCorr}) is the
sum of the positive contribution coming from the higher Fock states in the
pion, effectively taken into account by the second and third terms in (\ref%
{MPiGG}), and also the negative two-particle contribution due to the first
term in (\ref{MPiGG})\footnote{%
In \cite{AD0602} only part of the NLO power corrections has been discussed.}.
Note, that the model provides the opposite sign of the power correction
comparing with the QCD sum rule prediction \cite{svzB237}.

In general case at large $Q^{2}$ the model calculations reproduce the QCD
factorization result (\ref{AmplAsympt}),(\ref{J}) with the DAs given by
\begin{equation}
\varphi _{\pi }^{(2)}(x)=\frac{N_{c}}{4\pi ^{2}f_{\pi }^{2}}\int_{-\infty
}^{\infty }\frac{d\lambda }{2\pi }\int_{0}^{\infty }du\frac{F(u+i\lambda
\overline{x},u-i\lambda x)}{D\left( u-i\lambda x\right) D\left( u+i\lambda
\overline{x}\right) }\left[ xM\left( u+i\lambda \overline{x}\right) +\left(
x\leftrightarrow \overline{x}\right) \right] ,  \label{WF_VF2}
\end{equation}%
\begin{equation}
\varphi _{\pi }^{(4)}(x)=\frac{1}{\Delta ^{2}}\frac{N_{c}}{4\pi ^{2}f_{\pi
}^{2}}\int_{-\infty }^{\infty }\frac{d\lambda }{2\pi }\int_{0}^{\infty }du%
\frac{uF(u+i\lambda \overline{x},u-i\lambda x)}{D\left( u-i\lambda x\right)
D\left( u+i\lambda \overline{x}\right) }\left[ \overline{x}M\left(
u+i\lambda \overline{x}\right) +\left( x\leftrightarrow \overline{x}\right) %
\right] .  \label{WF_VF4}
\end{equation}%
In these expressions the $u$-variable plays the role of the quark transverse
momentum squared, $\overrightarrow{k}_{\perp }^{2}$, and $\lambda x,$ $%
-\lambda \overline{x}$ are the longitudinal projections of the quark
momentum on the light cone directions. The model DAs are defined at the
normalization scale characterized by the vacuum nonlocality $\mu ^{2}\sim
\Lambda ^{2}$. Concerning the LO DA, $\varphi _{\pi }^{(2)}(x)$, the
similar results within the instanton model have been earlier derived in \cite%
{WF2early}.

\begin{figure}[tbp]
\vskip -0.5cm \centering
\begin{minipage}[c]{7cm}
\epsfig{file=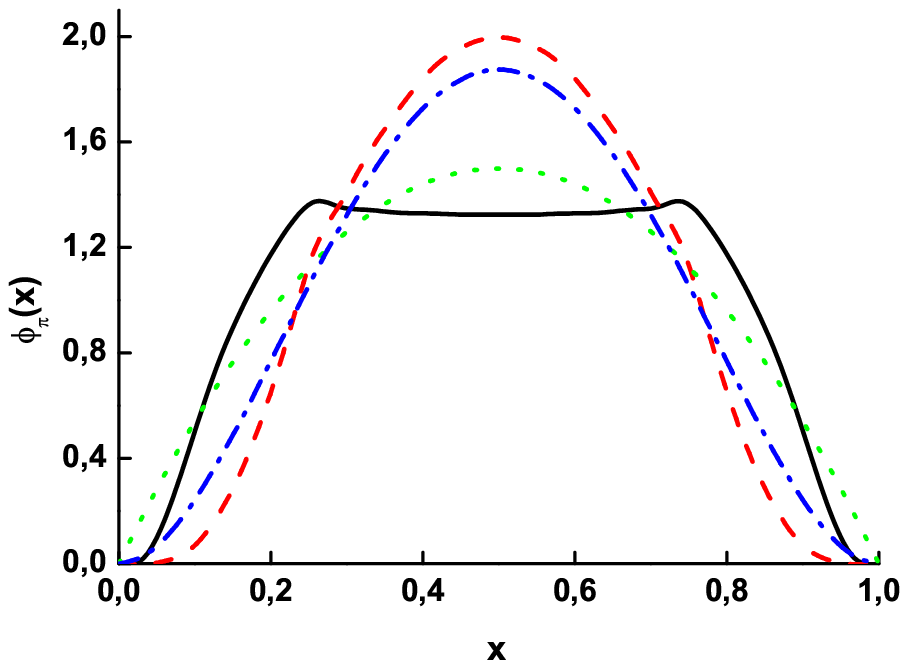,width=6.5cm,height=5.5cm}
\caption{\it  The pion distribution amplitudes (normalized by unity):
the model predictions for twist-2 (solid line) and twist-4 (dashed line) components
and the perturbative asymptotic limits of twist-2  (dotted line) and twist-4  (dash-dotted line) amplitudes.}
\label{f1}
\end{minipage}
\hspace*{0.5cm}
\begin{minipage}[c]{7cm}
\epsfig{file=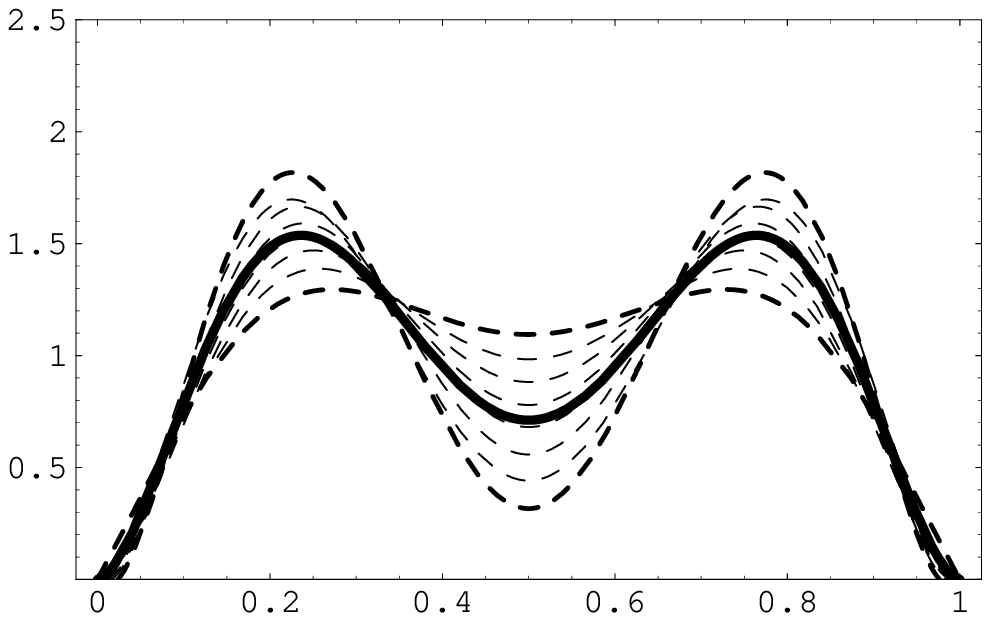,width=6.5cm,height=5.5cm}
\caption{\it An admisseble set of
the twist-2 pion distribution amplitudes (dashed lines, the best fit is solid line)
as predicted within
the QCD sum rules (from [15b]) with vacuum nonlocality parameter $\lambda_q^2 =0.4 GeV^2$
defined at $\mu^2 \approx 1 Gev^2$.}
\label{mb}
\end{minipage}
\end{figure}

In Fig. 1 the normalized by unity LO and NLO pion DAs are illustrated in
comparison with perturbative asymptotic DAs. For the numerical analysis the dynamical
mass profile is chosen in the Gaussian form $M(k^{2})=M_{q}\exp {%
(-2k^{2}/\Lambda ^{2})},$ where we take $M_{q}=350$ MeV and fix $\Lambda
=1.29$ GeV from the pion constant (\ref{f_pi}). Then, the value $\Delta
^{2}\equiv J^{(4)}\left( \omega =1\right) /$ $J^{(2)}\left( \omega =1\right)
=0.205$ GeV$^{2}$ is obtained which characterizes the scale of the power
corrections in the hard exclusive processes. The mean square radius of the
pion for the transition $\gamma ^{\ast }\pi ^{0}\rightarrow \gamma $ is $%
r_{\pi \gamma }^{2}=(0.566$ fm$)^{2}$ and numerically close to the value
derived from (\ref{Fpiggfit}). As it is clear from Fig. 1, the predicted
pion DAs at the realistic choice of the model parameters are close to the
asymptotic DAs. The corresponding conclusion with respect the LO DA is in
agreement with the results obtained in \cite{BakMikh,ADT99pigg} as it is seen
from comparison of Figs. 1 and 2.

The asymptotic coefficients $J^{(2,4)}(\omega )$ given by (\ref{J}), (\ref%
{WF_VF2}) and (\ref{WF_VF4}) may be identically rewritten in the form%
\begin{equation}
J^{(2)}\left( \omega \right) =-\frac{1}{\pi^2f_\pi}\int_{0}^{\infty
}duu\int_{0}^{\infty }dv\left\{ \frac{M^{1/2}\left( z_{-}\right) }{D(z_{-})}%
\frac{\partial }{\partial z_{+}}\left( \frac{M^{3/2}\left( z_{+}\right) }{%
D(z_{+})}\right) +\left( z_{-}\longleftrightarrow z_{+}\right) \right\} ,
\label{J2}
\end{equation}%
\begin{equation}
J^{(4)}\left( \omega \right) =\frac{2}{\pi^2f_\pi}\int_{0}^{\infty
}du\int_{0}^{\infty }dvv\left\{ \frac{M^{1/2}\left( z_{-}\right) }{D(z_{-})}%
\left[ \frac{M^{3/2}\left( z_{+}\right) }{D(z_{+})}+u\frac{\partial }{%
\partial z_{+}}\left( \frac{M^{3/2}\left( z_{+}\right) }{D(z_{+})}\right) %
\right] +\left( z_{-}\longleftrightarrow z_{+}\right) \right\} ,  \label{J4}
\end{equation}%
where $z_{\pm }=u+v(1\pm \omega )$. With the model parameters given above we
find the asymptotic coefficients $J^{(2)}\left( \omega =1\right) =0.171$ GeV
and $J^{(4)}\left( 1\right) /J^{(2)}\left( 1\right) =0.254$\ GeV$^{2}$ for
the process $\gamma \gamma ^{\ast }\rightarrow \pi ^{0}$. When the error in
the experimental fit is taken into account, the estimate of the LO
coefficient, $J^{(2)}\left( 1\right) ,$\ is in agreement with the fit
of CLEO data $J_{\exp }^{(2)}\left( 1\right) =0.16\pm 0.03$ GeV. The NLO power
correction, $\Delta ^{2}$, grows by 20\% with changing the kinematics from
equally distributed photon virtualities to asymmetric distribution.

\begin{figure}[tbp]
\vskip -0.5cm \centering
\begin{minipage}[c]{7cm}
\epsfig{file=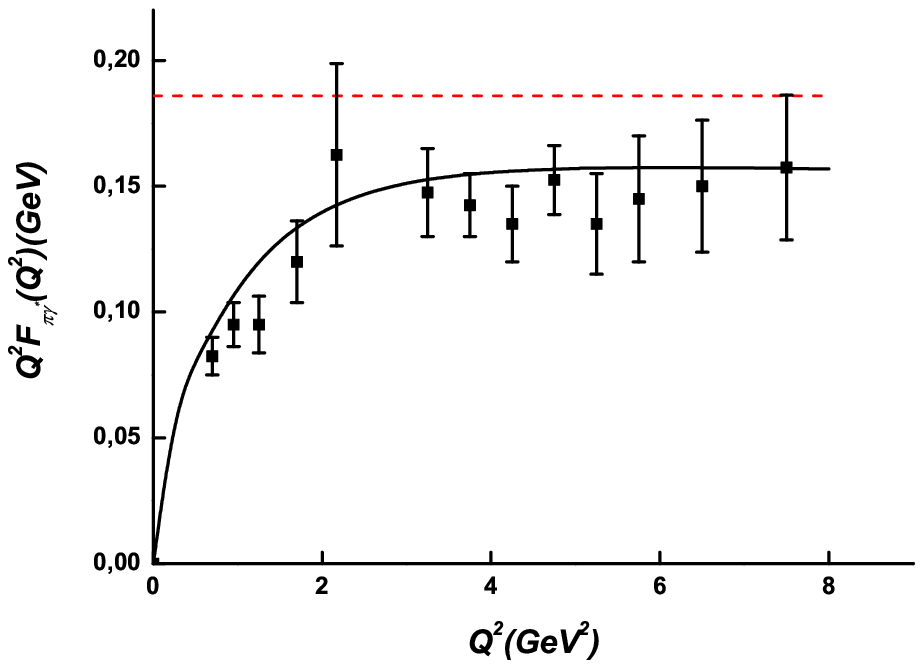,width=6.5cm,height=5.5cm}
\caption{\it  The pion-photon transition form factor $Q^2F_{\pi\gamma^\ast}(Q^2)$
 (solid line)
and its perturbative limit $2f_\pi$ (dotted line).
The experimental points ($Q^2F_{\pi\gamma^\ast\gamma}$) are taken from \cite{CLEO}.}
\label{f3}
\end{minipage}
\hspace*{0.5cm}
\begin{minipage}[c]{7cm}
\epsfig{file=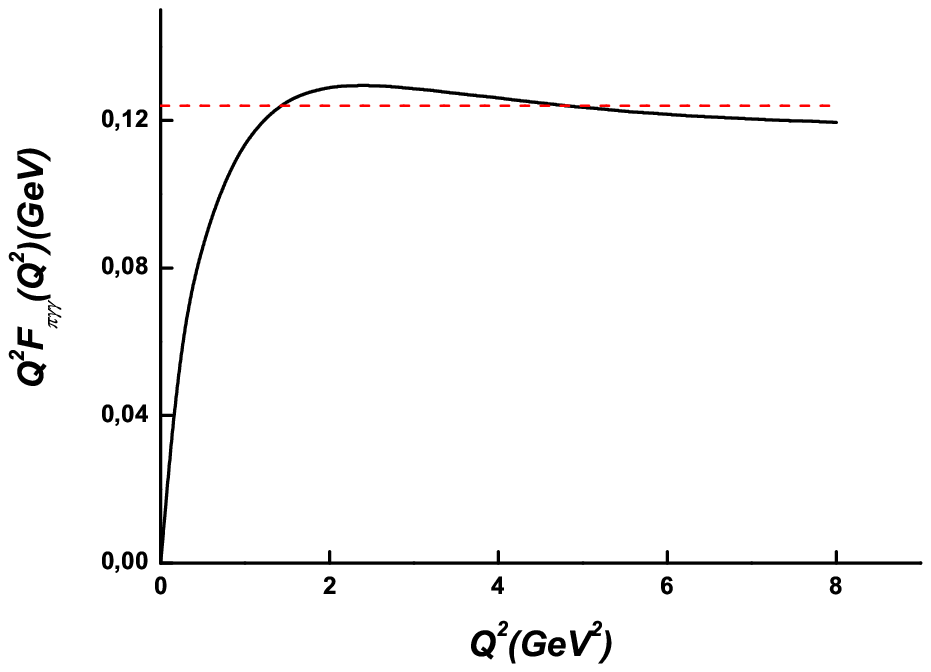,width=6.5cm,height=5.5cm}
\caption{\it The pion-photon transition form factor
$Q^2F_{\pi \gamma ^{\ast }\gamma ^{\ast}}(Q^{2})$
(solid line) and its perturbative limit $4f_\pi/3$ (dashed line).}
\label{f4}
\end{minipage}
\end{figure}

In Figs. 3 and 4 we plot the model predictions
for the form factors $F_{\pi \gamma\ast
}(Q^{2})=M_{\pi^{0}}\left( -Q^{2},0\right) $ and $F_{\pi \gamma ^{\ast
}\gamma ^{\ast}}(Q^{2})=M_{\pi ^{0}}\left( -Q^{2}/2,-Q^{2}/2\right)$
multiplied by square momentum $Q^{2}$ for the process $\gamma \gamma ^{\ast
}\rightarrow \pi ^{0}$ and $\gamma ^{\ast }\gamma ^{\ast }\rightarrow \pi
^{0}$, correspondingly. In Fig. 3 we also indicate the CLEO data.
In the model form factors the perturbative $\alpha
_{s}-$ corrections \cite{Braaten} to the leading twist-2 term are taken into
account with the running coupling, $\alpha _{s}(Q^{2})$, that has zero at
zero momentum \cite{VAV02}. With such effective behaviour in the infrared
region the perturbative corrections do not influent the chiral anomaly. At
high momentum squared the leading perturbative correction provides negative
contribution to the form factors and compensate the NLO power corrections in
the region $2-10$ GeV$^{2}$. The unknown perturbative corrections to the
twist-4 contribution is considered as inessential. The power corrections
generated by the twist-3 pion DAs are also negligible since they are
proportional to the small current quark mass.

In conclusion, within the covariant nonlocal model describing the quark-pion
dynamics, we obtain the $\pi \gamma ^{\ast }\gamma ^{\ast }$ transition form
factor in the region up to moderately high momentum transfer squared, where
the perturbative QCD evolution does not reach the asymptotic regime yet.
From the comparison of the kinematical dependence of the
coefficients of the power expansion in $1/Q^2$ of the transition pion form factor,
as it is given by pQCD and
the nonperturbative model, the relations Eqs. (\ref{WF_VF2}, \ref{WF_VF4})
between the pion DAs and the dynamical quark mass and quark-pion vertex are
derived. The other possible sources of contributions to the form factor
arise from inclusion into the model of the low lying vector and axial-vector
mesons. They do not change the result given by the chiral anomaly (\ref{ChAn}%
) for the two-gamma pion decay. The contributions of the vector mesons to
the leading order asymptotics of the form factor are expected to be small,
but they may be more important in treating the twist-4 power corrections and
the pion mean radius.

{\it Acknowledgments}

I am grateful to A.P. Bakulev, W. Broniowski, A. Di Giacomo, A.S. Gorski,
N.I. Kochelev, S.V. Mikhailov, M.K. Volkov, L. Tomio and V.L. Yudichev for
many useful discussions on topics related to this work. The work is
supported by RFBR Grants 01-02-16431 and 02-02-16194 and by Grant
INTAS-2000-366.

\end{document}